# Spin glass anomalies in stoichiometric spin-chain oxides, $Ca_3CoXO_6$ (X = Rh, Ir and Co)

S. RAYAPROL, KAUSIK SENGUPTA AND E.V. SAMPATHKUMARAN
Tata Institute of Fundamental Research, Homi Bhabha Road, Mumbai – 400 005

*Abstract*

*The results of ac susceptibility ($\chi$) measurements for $Ca_3CoXO_6$ (X = Rh, Ir and Co) compounds, containing spin-chains separated by Ca ions, are presented. All these compounds exhibit an unusually large frequency dependence of ac $\chi$ in the vicinity of respective magnetic ordering temperatures, which is normally not encountered in conventional spin-glasses. The frequency dependence of ac $\chi$ peak temperature is however found to obey Vogel-Fulcher relationship. The results thus establish that these compounds are unusual spin-glasses, that too among stoichiometric compounds presumably due to geometrical frustration.*

## INTRODUCTION

There is a considerable interest in the current literature to understand the properties of quasi-one-dimensional magnetic systems. In this respect, we have undertaken a systematic investigation [1-4] of the compounds of the type $A_3MXO_6$, which are derived [5-7] from $K_4CdCl_6$ type rhombohedral structure (space group, $R\bar{3}c$). In these systems, usually 'A' is either Ca or Sr, 'M' is a magnetic or non-magnetic metal ion (e.g., Cu, Co, Zn, etc) and X is usually a magnetic metal ion (e.g., Co, Mn, Ir, Pt etc). The $MO_3$ trigonal prisms and $XO_3$ octahedra share a face along the c-axis and are placed alternatively forming a chain along the c-axis. These chains are separated from each other by 'A' ions. The relative strengths of intrachain and interchain magnetic couplings, and hence quasi-one-dimensional nature, can be controlled by varying M and X ions. In addition, the triangular arrangement of the magnetic ions in the basal plane is expected to lead to magnetic frustration in the event that interchain interaction is antiferromagnetic. The results available till todate indicate that some of the compounds, viz., $Ca_3CoXO_6$ (X = Rh, Ir and Co), indeed exhibit complex magnetic behavior [2-4, 7-9] arising from both Co and X ions, which are yet to be understood completely. In this article, we focus our attention on ac susceptibility ($\chi$) behavior to establish that these compounds are indeed spin-glasses of an unusual type.

## EXPERIMENTAL

The polycrystalline samples of the $Ca_3CoXO_6$ (X = Rh, Ir and Co) series were synthesized by solid state reaction method. The appropriate quantities of the starting compounds - $CaCO_3$, $Co_3O_4$, Rh metal powder and Ir metal powder (> 99.9 % pure) - were ground thoroughly under acetone. The samples were then calcined in the temperature range 900 – 1000 $^0C$ and sintered (after palletizing) in the temperature range of 1100 – 1200 $^0C$. The samples were characterized by x-ray diffraction (Cu-$K_\alpha$) and found to be single phase, with all forming in the rhombohedral structure. The lattice constants are in good agreement with the literature values. The ac $\chi$ measurements down to 1.8 K at various frequencies ($\nu$) were carried out (ac field= 1 Oe) employing a commercial magnetometer (in zero dc magnetic field).

## RESULTS AND DISCUSSION

Figure 1(a & b) shows the real and imaginary parts of ac $\chi$ for all the three samples obtained at different $\nu$.

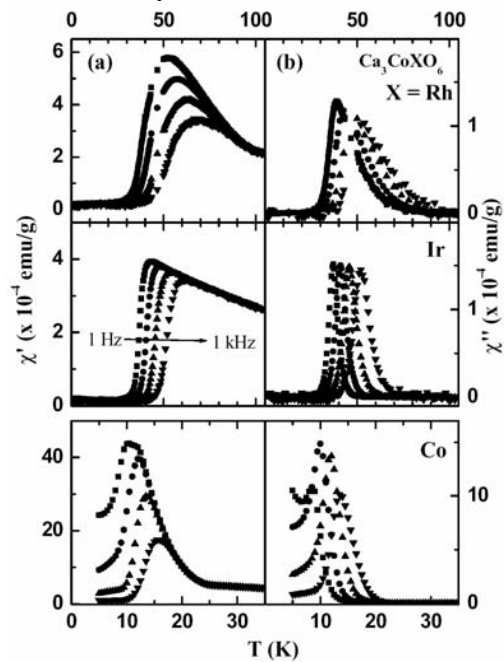

**Fig 1** (a) The real part and (b) imaginary parts of ac $\chi$ as a function of temperature for all the three samples. The plots shift to the higher temperature side with increasing $\nu$. $\nu$ = 1, 10, 100 and 1000 Hz.

There is a distinct peak in all cases, establishing the existence of magnetic ordering for all cases. There is an upward shift of the peak with increasing ν, as though these compounds could be classified as spin-glasses. However, what is puzzling is that the observed magnitude of the shift in the peak temperature is much larger than what is known for conventional spin-glasses. Thus, for instance, for X= Ir, the curve shifts as much as 15 K for a change of ν from 1 Hz to 1 kHz, yielding a value close to 0.1 for the factor $\Delta T_f/T_F\Delta(\ln \nu)$ (Ref. 10), a value much larger than that known for conventional spin-glasses. This raises a question whether these compounds can be classified as spin-glasses.

In order to address this question, we have analyzed the data in terms of Voger-Fulcher relationship – a well-known testing ground for spin-glass phenomenon. The relationship is:

$$\nu = \nu_0 \exp[-E_a/k_B(T_f-T_a)] \qquad --- (1)$$

where ν is the driving frequency of our ac χ measurement, $\nu_0$ = ideal spin glass frequency ~ $10^8$ Hz, $T_f$ = peak temperature and $T_a$ = ideal glass temperature, which is believed to be a measure of interaction strengths between clusters in the spin glass phase. Therefore, $T_f$ should vary linearly with $1/\ln(\nu_0/\nu)$ for spin-glasses. It is indeed found to be the case in all our compounds as shown in figure 2 thus establishing the validity of Voger-Fulcher relationship for all these cases. The parameters obtained by such an analysis are plotted in Fig 3. In order to see whether the coupling strengths in some way determined by c/a ratio of the unit-cell have no any role to play on the parameters, we have also shown corresponding c/a values. Clearly, these parameters do not seem to follow c/a as one varies X (Rh → Ir → Co).

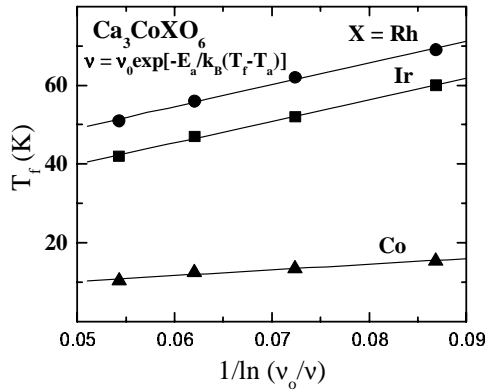

**Fig 2** The plot to show the validity of Voger-Fulcher relationship for all the three samples marked in the figure. For each compound, the line passing through the data points is obtained by a linear fit.

## CONCLUSION

There is a dramatically large frequency dependence of ac susceptibility in the vicinity of respective magnetic ordering temperatures for the spin-chain $Ca_3CoXO_6$ compounds (X = Rh, Ir and Co). Such an anomaly has, to our knowledge been not been known among stoichiometric magnetic materials. Interestingly, Voger-Fulcher relationship is found to be applicable, thereby establishing that these compounds can be classified as spin-glasses. Thus, these materials serve as examples for spin-glass behavior among stoichiometric compounds presumably due to topological frustration effects. Finally, we would like to mention a fascinating observation we made [4]: creation of disorder by chemical substitution restores long range magnetic ordering (exactly opposite to what has been known in magnetism till todate), as though the geometrical magnetic frustration is released by disorder. Thus, these materials reveal a wealth of new information in magnetism and therefore are expected to attract further attention of the community.

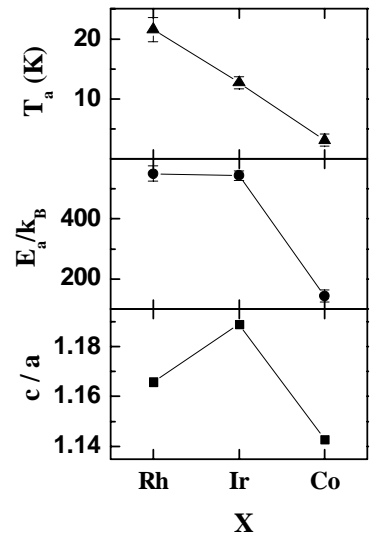

**Fig 3** Values of ideal spin glass temperature ($T_a$), $E_a/k_B$ and ratio of lattice constants (c /a) for all the three samples.

## REFERENCES

1. See, for instance, Asad Niazi et al. Phys. Rev. Lett. **88** (2002) 107202; M. Loewenhaupt et al. Europhys. Let. **63** (2003) 374; Kausik Sengupta et al. Phys. Rev. B **68** (2003) 012411 and references therein
2. E. V. Sampathkumaran et al. Phys. Rev. **B 65** (2002) 180401R
3. S. Rayaprol et al. Phys. Rev. **B 67** (2003) 180404R
4. S. Rayaprol et al. Solid State. Commun. **128** (2003) 79
5. T. N. Nguyen et al. J. Solid State Chem. **117** (1995) 300
6. H. Kageyama et al. J. Solid State Chem. **140** (1998) 14
7. S. Niitaka et al. J. Solid State Chem. **146** (1999) 137; S. Niitaka et al Phys. Rev. Lett. **87** (2001) 177202
8. H. Kageyama et al. J. Phys. Soc. Jpn. **66** (1997) 1607
9. A. Maignon et al. Eur. Phys. J. B **15** (2000) 657
10. K. Binder et al. Rev. Mod. Phys. **58** (1986) 801